\documentclass[11pt]{article}
\usepackage{epsfig}
\usepackage{hyperref}
\usepackage{url}
\usepackage{graphicx}
\usepackage{dcolumn}
\usepackage{bm}
\usepackage{graphicx,amssymb}
\usepackage{color}
\usepackage{float}
\newcommand{\pt}{$p_{ {\mathrm T} }$}
\newcommand{\y}{$y$}

\usepackage{multirow}
\usepackage{rotating}
\usepackage{lineno}
\usepackage{subfigure}
\usepackage{lineno}
\begin{document}


\begin{titlepage}

\title{\Large \bf Inclusive J/$\psi$ and $\psi$(2S) production in pp and p-Pb collisions at forward rapidity with ALICE at the LHC}
\author{\large Biswarup Paul (for the ALICE Collaboration)\\
\textit {Saha Institute of Nuclear Physics, Kolkata - 700064, India}}

\date{}

\maketitle

The ALICE collaboration has studied inclusive J/$\psi$ and $\psi$(2S) production at forward rapidities in pp collisions at $\sqrt{s} = 7$ TeV with the ALICE Muon Spectrometer. The analysis has been carried out on a data sample corresponding to an integrated luminosity $\mathcal{L}_{\rm int}$ =
1.35 pb$^{-1}$.
The production cross-sections of J/$\psi$ and $\psi$(2S), integrated over the transverse momentum (0 $<$ $p_{ {\mathrm T} }$ $<$ 20 GeV/$c$) and rapidity (2.5 $<$ $y$ $<$ 4), have been measured. The J/$\psi$ and $\psi$(2S) differential cross-sections, in transverse momentum and rapidity, have also been measured, significantly extending the $p_{ {\mathrm T} }$ reach of previous measurements performed in the same $y$-range.
The results have been compared with the previously published ALICE results ($\mathcal{L}_{\rm int}$ = 15.6 nb$^{-1}$) and also with the measurement performed by the LHCb collaboration. The $\psi$(2S)/J/$\psi$ ratio, integrated over $p_{ {\mathrm T} }$ and $y$, has been measured.
This ratio has also been evaluated as a function of transverse momentum and rapidity and compared with the LHCb measurement. Finally, recent results on cross-sections, $\psi$(2S)/J/$\psi$ production ratio, nuclear modification factor ($R_{\rm pPb}$) and forward-to-backward yield ratio ($R_{\rm FB}$) in p-Pb collisions at $\sqrt{s_{\rm NN}} = 5.02$ TeV will be discussed.
\end{titlepage}

\section{Introduction}
Fourty years have passed after the discovery of charmonia (bound states of $c$ and $\overline c$ quarks) but the understanding of their production in hadronic collisions still remains incomplete. Models based on Quantum Chromodynamics (QCD) provide a qualitative understanding of their production mechanism in pp collisions at high \pt\ but are not suitable for the description of the low-\pt\ region. The ALICE collaboration has measured the production of charmonia down to \pt=0. ALICE results are thus useful for understanding the low-\pt\ production mechanisms. In addition, the pp results for the charmonium provide a baseline for the nuclear modification factor of charmonium production in \mbox{p-Pb} and \mbox{Pb-Pb} collisions. The study of charmonia in \mbox{p-Pb} collisions can be used as a tool for a quantitative investigation of the cold nuclear matter (CNM) effects and various mechanisms such as nuclear parton shadowing, $c\overline c$ break-up via interaction with nucleons, initial/final state energy loss, relevant in the context of studies of the strong interaction. Strong shadowing and coherent energy loss effects are expected at the LHC.

The Muon Spectrometer of ALICE~\cite{R1} is designed to measure the charmonium (J/$\psi$,$\psi(2\rm S)$) and bottomonium ($\Upsilon(1\rm S)$,$\Upsilon(2\rm S)$,$\Upsilon(3\rm S)$) states in the pseudo-rapidity interval $-$4 $<$ $\eta$ $<$ $-$2.5.

\section{pp}
\subsection{Event and track selections}
The analysis described in this document (see~\cite{R5} for details) uses pp collision data at $\sqrt{s} = 7$ TeV. The data were recorded in 2011 with a trigger defined by the coincidence of a minimum bias trigger with the detection of two opposite sign muons reconstructed in the trigger chambers of the muon spectrometer. A total of 4 million events were analyzed, corresponding to an integrated luminosity $\mathcal{L}_{\rm int}$ = 1.35 pb$^{-1}$ (with 5\% systematic uncertainty). In order to improve the purity of the muon tracks the following selection criteria were applied: 
(1) both muon tracks matched with trigger tracks above 1 GeV/$\it{c}$ \pt\ threshold,
(2) both muon tracks in the pseudo-rapidity range $-$4 $<$ $\eta$ $<$ $-$2.5,
(3) transverse radius coordinate of the tracks at the end of the hadron absorber (the longitudinal position of the absorber from the interation point (IP) is $-$5.0 $<$ $z$ $<$ $-$0.9 m) in the range 17.6 $<$ $R_{\rm{abs}}$ $<$ 89.5 cm,
(4) dimuon rapidity in the range  2.5 $<$ \y\ $<$ 4.0 (a positive rapidity value is chosen by convention) and 
(5) dimuon \pt\ in the range  0 $<$ \pt\ $<$ 20 GeV/$\it{c}$.

\subsection{Signal extraction}

\begin{figure}[H]
     \begin{center}
\includegraphics[scale=0.45]{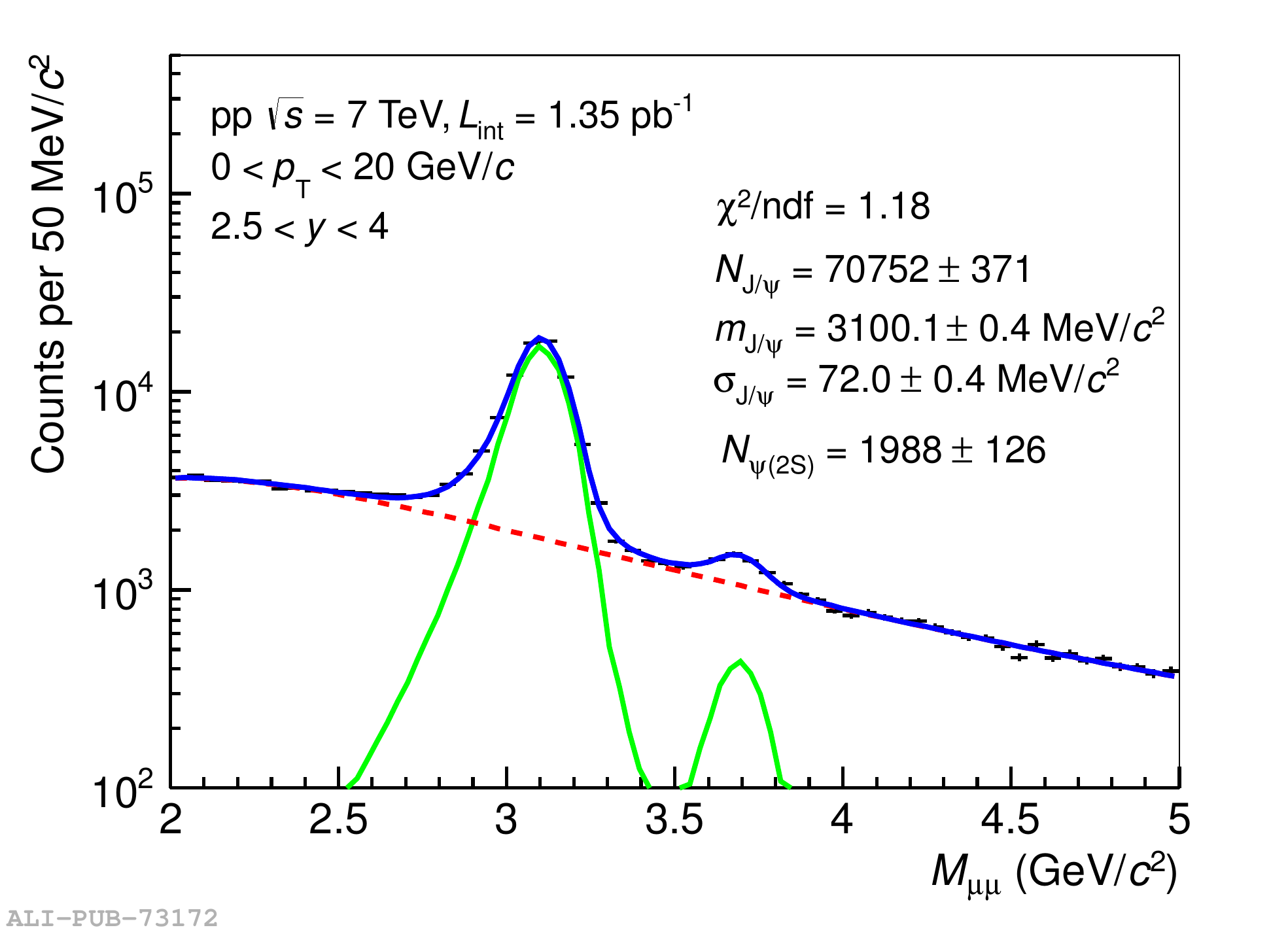}
\vspace{-2.00mm}
\caption{\label{fig1} Opposite sign dimuon invariant mass distribution, integrated over \y\ (2.5 $<$ \y\ $<$ 4.0) and \pt\ (0 $<$ \pt\ $<$ 20 GeV/$\it{c}$).}
     \end{center}
\end{figure}

Fig.~\ref{fig1} shows the dimuon invariant mass distribution fitted with two Extended Crystal Ball functions (Crystal Ball function with a non Gaussian tail on both sides) for both the J/$\psi$ and the $\psi$(2S) and a variable width Gaussian function (a Gaussian function with a width varying linearly with the mass) for the background.

\subsection{Integrated and differential production cross-sections of J/$\psi$ and $\psi$(2S)}
The production cross-sections of J/$\psi$ and $\psi$(2S) have been determined as:
\begin{eqnarray}
  \sigma_{\psi} = \frac{N_{\psi}}{(A {\times \epsilon})}.\frac{1}{\rm B.R._{\psi \rightarrow \mu^{+}\mu^{-}}}.\frac{1}{\mathcal{L}_{\rm int}}
\label{eq:PsiPsigma}
\end{eqnarray}
where $N_{\psi}$ is the number of J/$\psi$ or $\psi$(2S) obtained from the signal extraction,
$A$$\times \epsilon$ is the acceptance times efficiency correction factor of J/$\psi$ or $\psi$(2S),
B.R.$_{\psi \rightarrow \mu^{+}\mu^{-}}$ is the branching ratio of J/$\psi$ or $\psi$(2S) in the dimuon decay channel (B.R.$_{J/\psi \rightarrow \mu^{+}\mu^{-}}$ = (5.93$\pm$0.06)\% and B.R.$_{\psi(2S) \rightarrow \mu^{+}\mu^{-}}$ = (0.78$\pm$0.09)\%) and
$\mathcal{L}_{\rm int}$ is the integrated luminosity.

\begin{figure}[H]
     \begin{center}
        \subfigure[\hspace{0.2cm}\pt\ differential cross-section of J/$\psi$.]{%
            \label{fig2a}
            \includegraphics[width=0.49\textwidth]{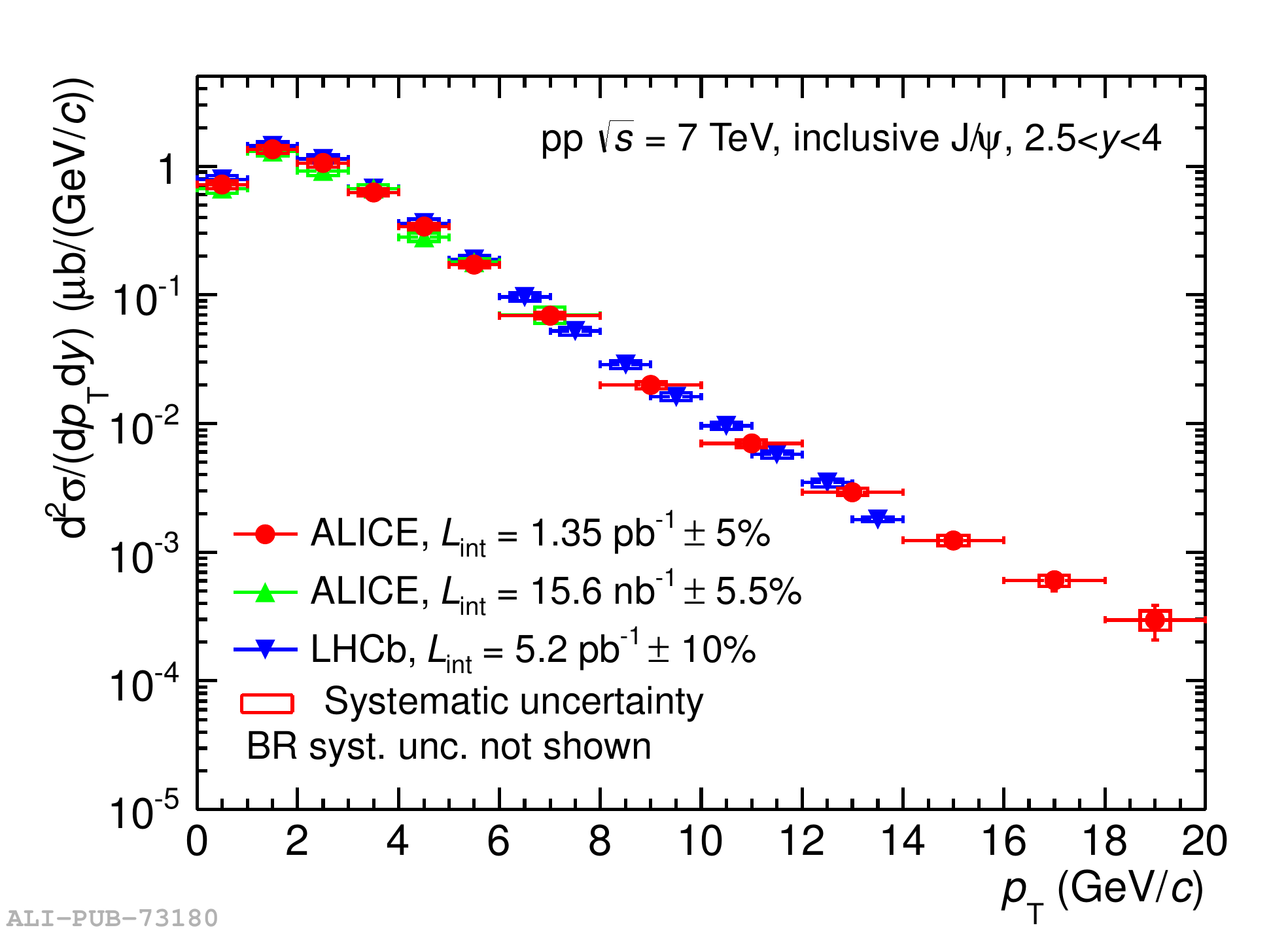}
        }%
        \subfigure[\hspace{0.2cm}\y\ differential cross-section of J/$\psi$.]{%
           \label{fig2b}
           \includegraphics[width=0.49\textwidth]{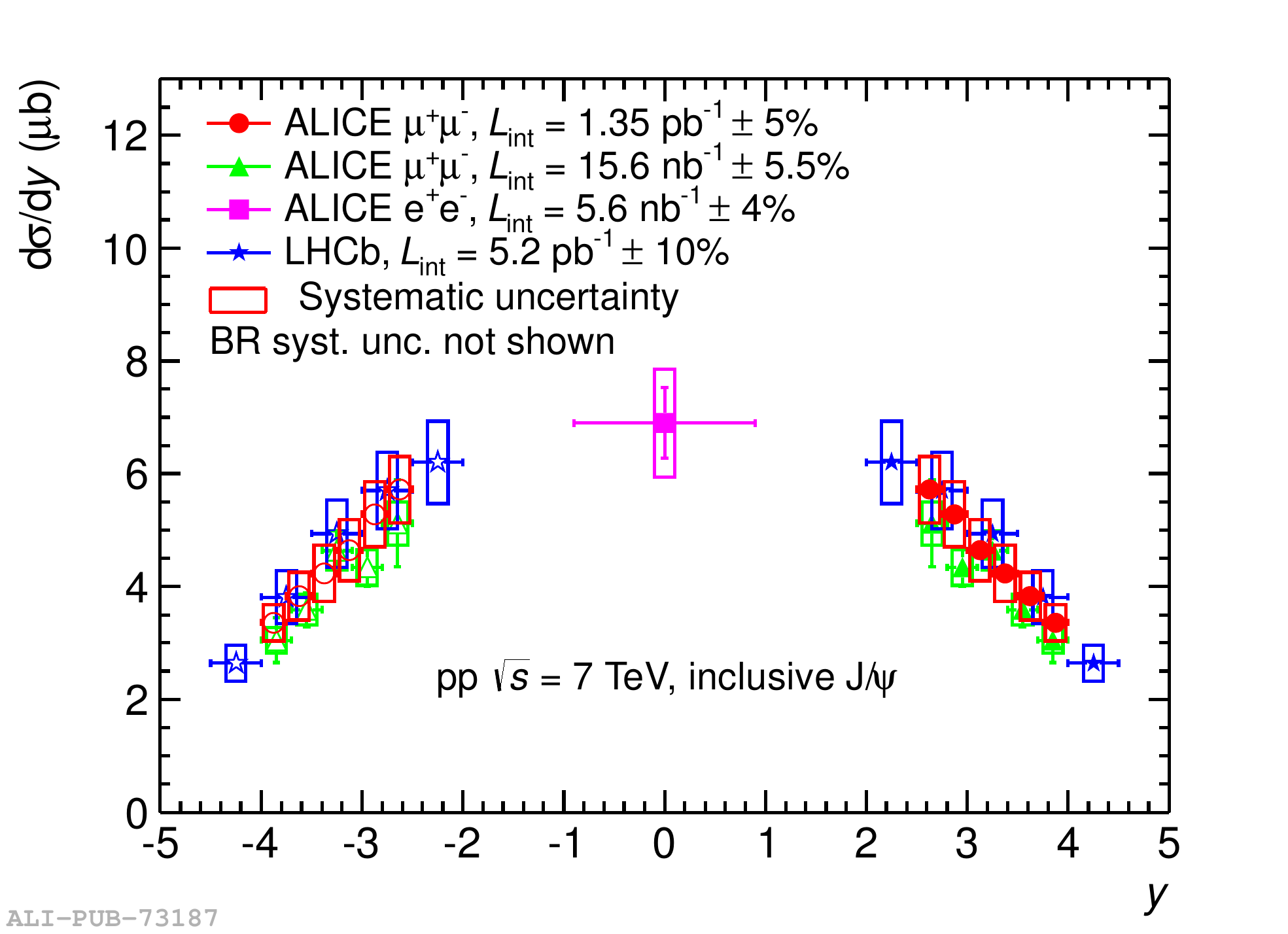}
        } 
        \subfigure[\hspace{0.2cm}\pt\ differential cross-section of $\psi$(2S).]{%
            \label{fig2c}
            \includegraphics[width=0.49\textwidth]{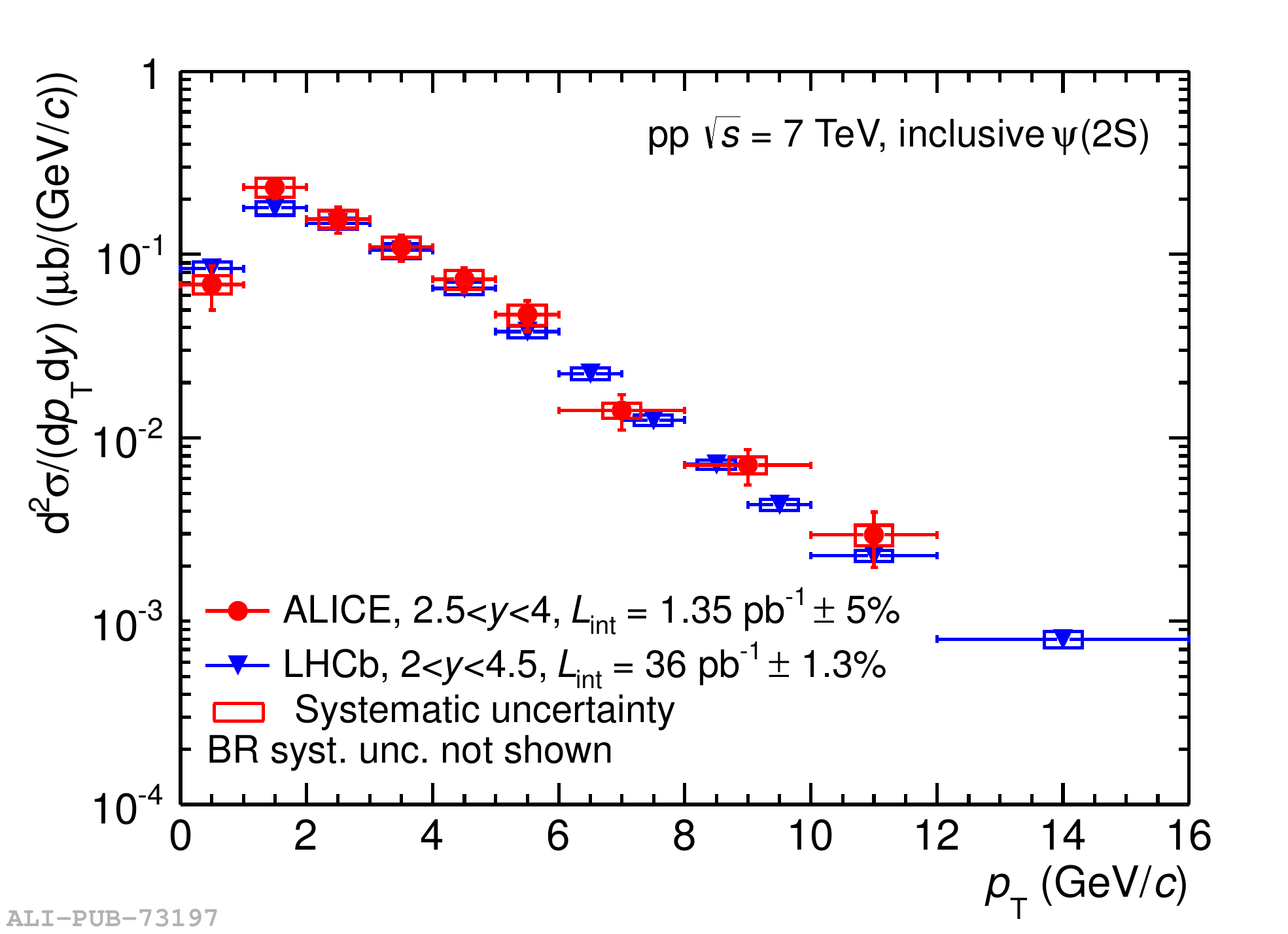}
        }%
        \subfigure[\hspace{0.2cm}\y\ differential cross-section of $\psi$(2S).]{%
           \label{fig2d}
           \includegraphics[width=0.49\textwidth]{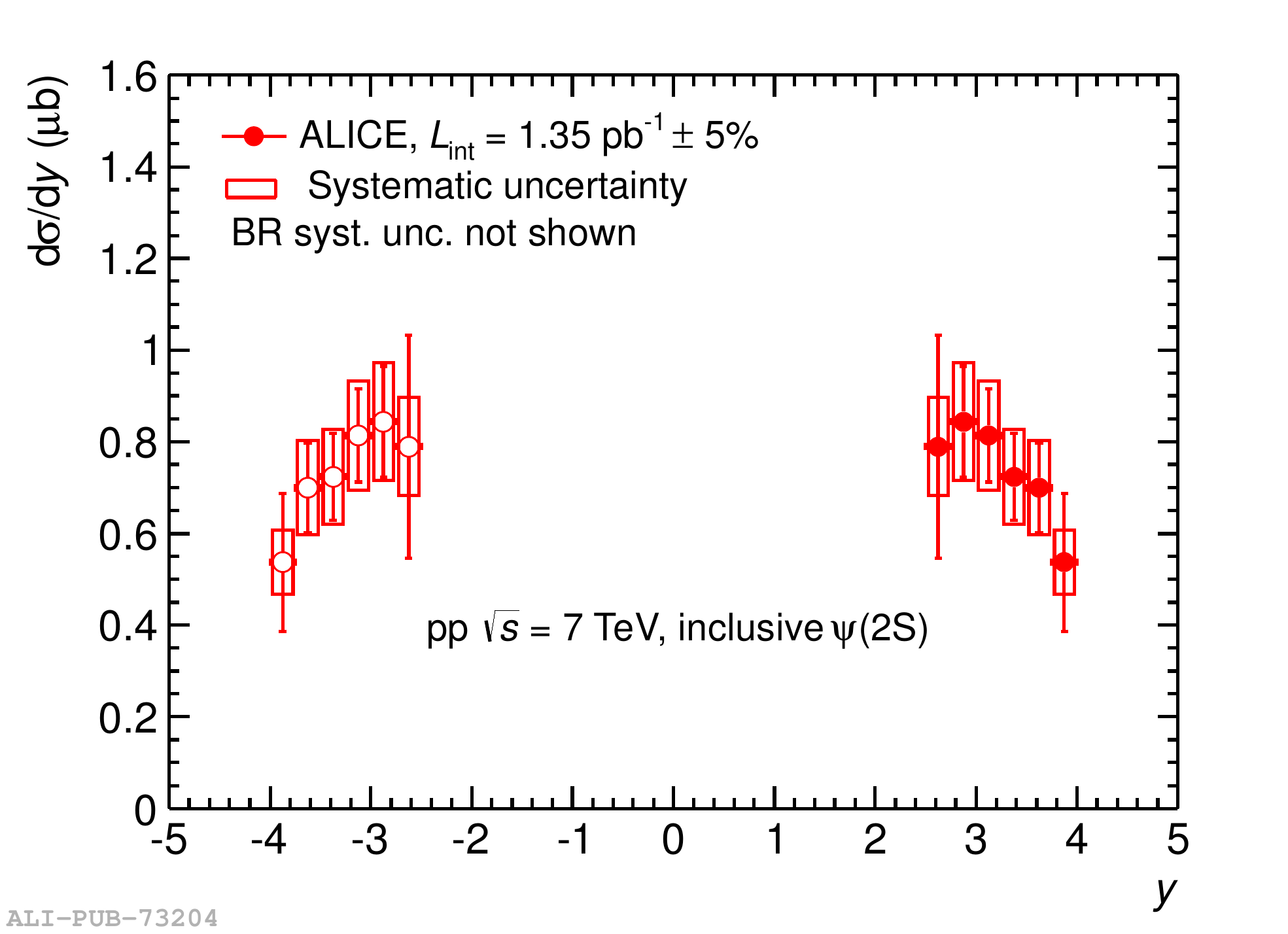}
        } 
    \end{center}
    \caption{%
Differential cross-sections of J/$\psi$ and $\psi$(2S).
     }%
   \label{fig2}
\end{figure}

The measured production cross-sections of J/$\psi$ and $\psi$(2S), integrated over the \y\ and \pt\ ranges, are:
\begin{center}
$\sigma_{\rm J/\psi}$~=~6.69~$\pm$~0.04~(stat.)~$\pm$~0.63~(syst.)~$\mu$b.\\
$\sigma_{\rm \psi(2S)}$~=~1.13~$\pm$~0.07~(stat.)~$\pm$~0.19~(syst.)~$\mu$b.
\end{center}
Fig.~\ref{fig2a} and Fig.~\ref{fig2b} show the differential production cross-sections of J/$\psi$ as a function of \pt\ and \y, respectively. This result is consistent with the previous ALICE result~\cite{R2} and also with the measurement performed by the LHCb collaboration~\cite{R3}. This measurement extends the \pt\ reach to 20 GeV/$\it{c}$ at forward rapidity.

Fig.~\ref{fig2c} shows the differential production cross-sections of $\psi$(2S) as a function of \pt. The result is consistent with the LHCb measurement~\cite{R4} in the same rapidity interval. Fig.~\ref{fig2d} shows the differential production cross-sections of the $\psi$(2S) as a function of \y. This is the first measurement of $\psi$(2S) differential cross-sections as a function of \y\ in pp collisions at $\sqrt{s} = 7$ TeV. 

\subsection{$\psi$(2S) to J/$\psi$ cross-section ratio}

\begin{figure}[H]
     \begin{center}
        \subfigure[\hspace{0.2cm}$\psi$(2S) to J/$\psi$ cross-section ratio vs \pt.]{%
            \label{fig3a}
            \includegraphics[width=0.49\textwidth]{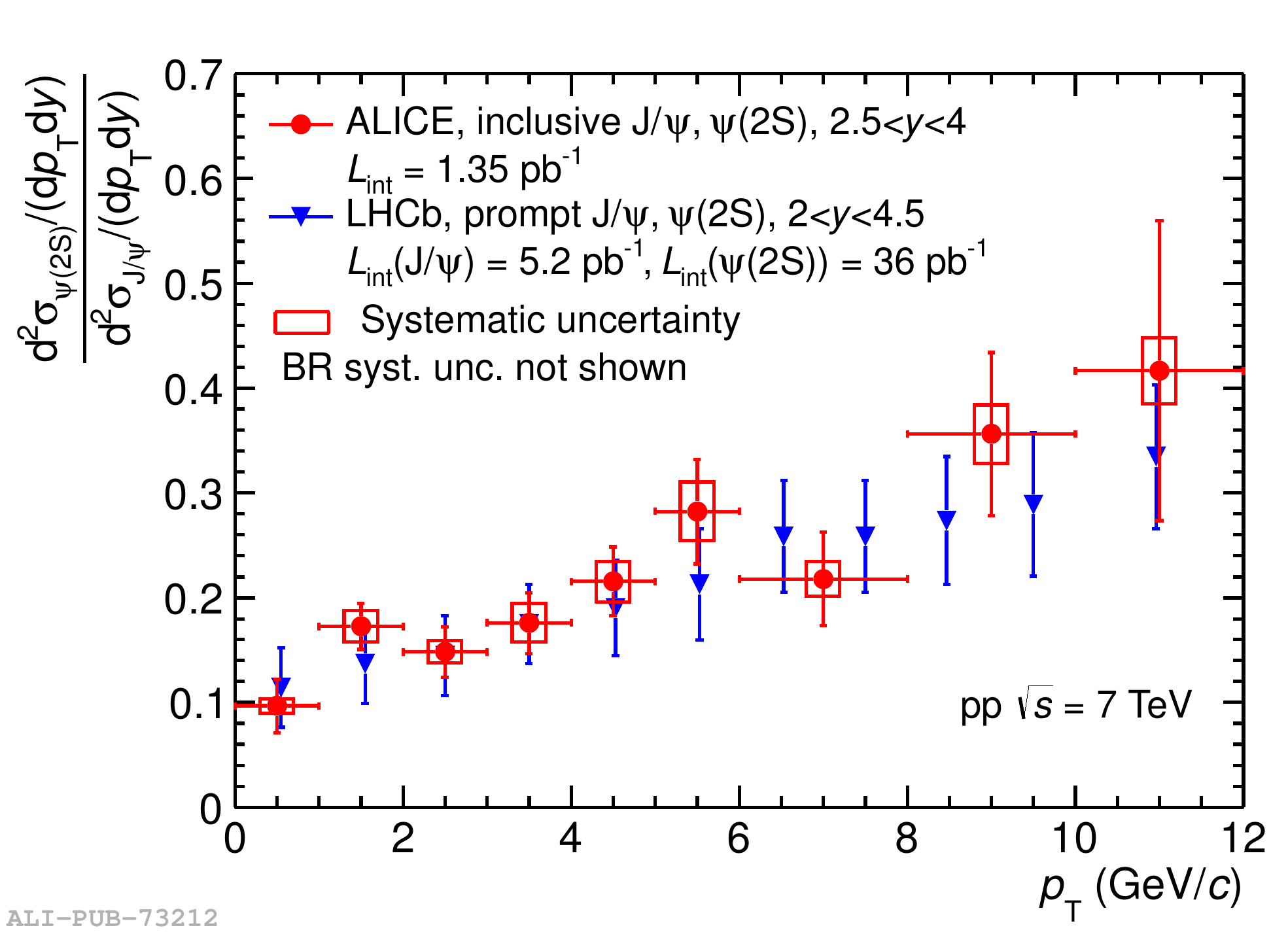}
        }%
        \subfigure[\hspace{0.2cm}$\psi$(2S) to J/$\psi$ cross-section ratio vs \y.]{%
           \label{fig3b}
           \includegraphics[width=0.49\textwidth]{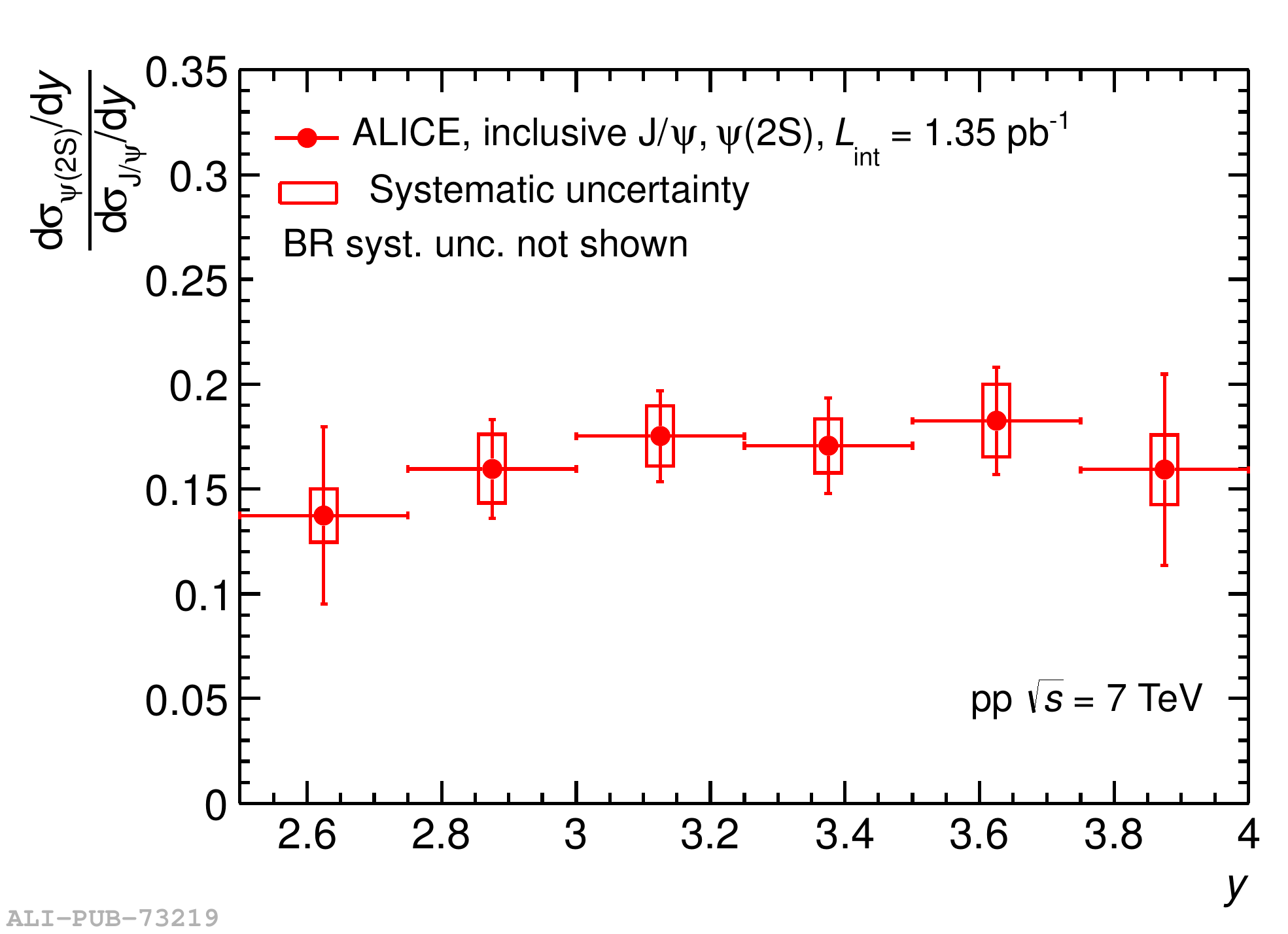}
        } 
    \end{center}
    \caption{%
$\psi$(2S) to J/$\psi$ cross-section ratio.
     }%
   \label{fig3}
\end{figure}

The measured inclusive $\psi$(2S) to J/$\psi$ cross-section ratio, integrated over \pt\ and \y, is 0.170 $\pm$ 0.011 (stat.) $\pm$ 0.013 (syst.). This ratio was also measured as a function of \pt\ and \y\ as shown in Fig.~\ref{fig3a} and Fig.~\ref{fig3b}, respectively. A clear \pt\ dependence can be observed, consistent with the one measured by LHCb~\cite{R4}. No strong \y\ dependence is visible, in the \y\ range covered by the ALICE muon spectrometer. More details of pp analysis are avaliable in~\cite{R5}. 

\section{p-Pb}
The two-in-one magnet design of the LHC imposes the same magnetic rigidity of the beams in the two rings. In p-Pb collisions, beam 1 consisted of protons at 4 TeV energy circulating towards the Muon spectrometer in the negative z direction while beam 2 consisted of fully stripped Pb ions at 82$\times$4 TeV energy circulating in the positive z direction. This configuration resulted in collisions at $\sqrt{s}_{\rm NN} = 5.02$ TeV and in a nucleon-nucleon center-of-mass system that moves with a rapidity $\Delta{y}_{\rm NN}$ = 0.465 in the proton beam direction. In the center-of-mass system, the Muon spectrometer covers the region 2.03 $<$ \y$_{\rm cms}$ $<$ 3.53 in p-Pb collisions. By inverting the directions of the two beams (Pb-p configuration) the region $-$4.46 $<$ \y$_{\rm cms}$ $<$ $-$2.96 has been also covered.

\subsection{$\psi$(2S)/J/$\psi$ ratio}
The ratio B.R.$_{\rm \psi(2S)\rightarrow \mu^{+}\mu^{-}}$$\times$$\sigma_{\rm \psi(2S)}$/B.R.$_{\rm J/\psi \rightarrow \mu^{+}\mu^{-}}$$\times$$\sigma_{\rm J/\psi}$ (i.e. the $\psi$(2S)/J/$\psi$ ratio not corrected for branching ratios of J/$\psi$ and $\psi$(2S)) for p-Pb and Pb-p, integrated over \pt\ and \y\ are 0.0154 $\pm$ 0.0019 (stat.) $\pm$ 0.0015 (syst.) and 0.0116 $\pm$ 0.0018 (stat.) $\pm$ 0.0011 (syst.), respectively. 

\begin{figure}[H]
     \begin{center}
        \subfigure[\hspace{0.2cm}$\psi$(2S)/J/$\psi$ ratio for pp and p-Pb.]{%
            \label{fig4a}
            \includegraphics[width=0.49\textwidth]{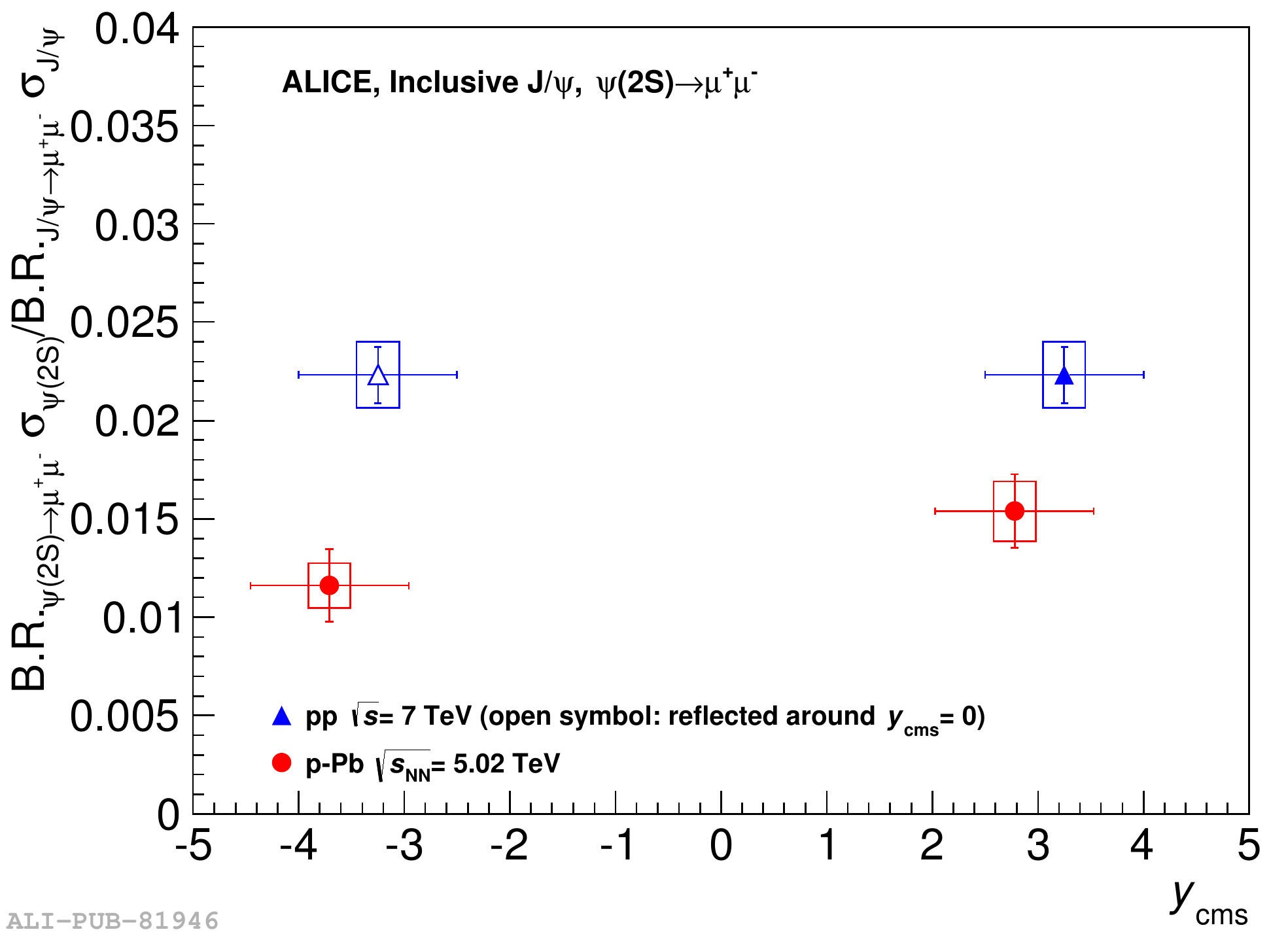}
        }%
        \subfigure[\hspace{0.2cm}Double ratio for p-Pb collisions.]{%
           \label{fig4b}
           \includegraphics[width=0.49\textwidth]{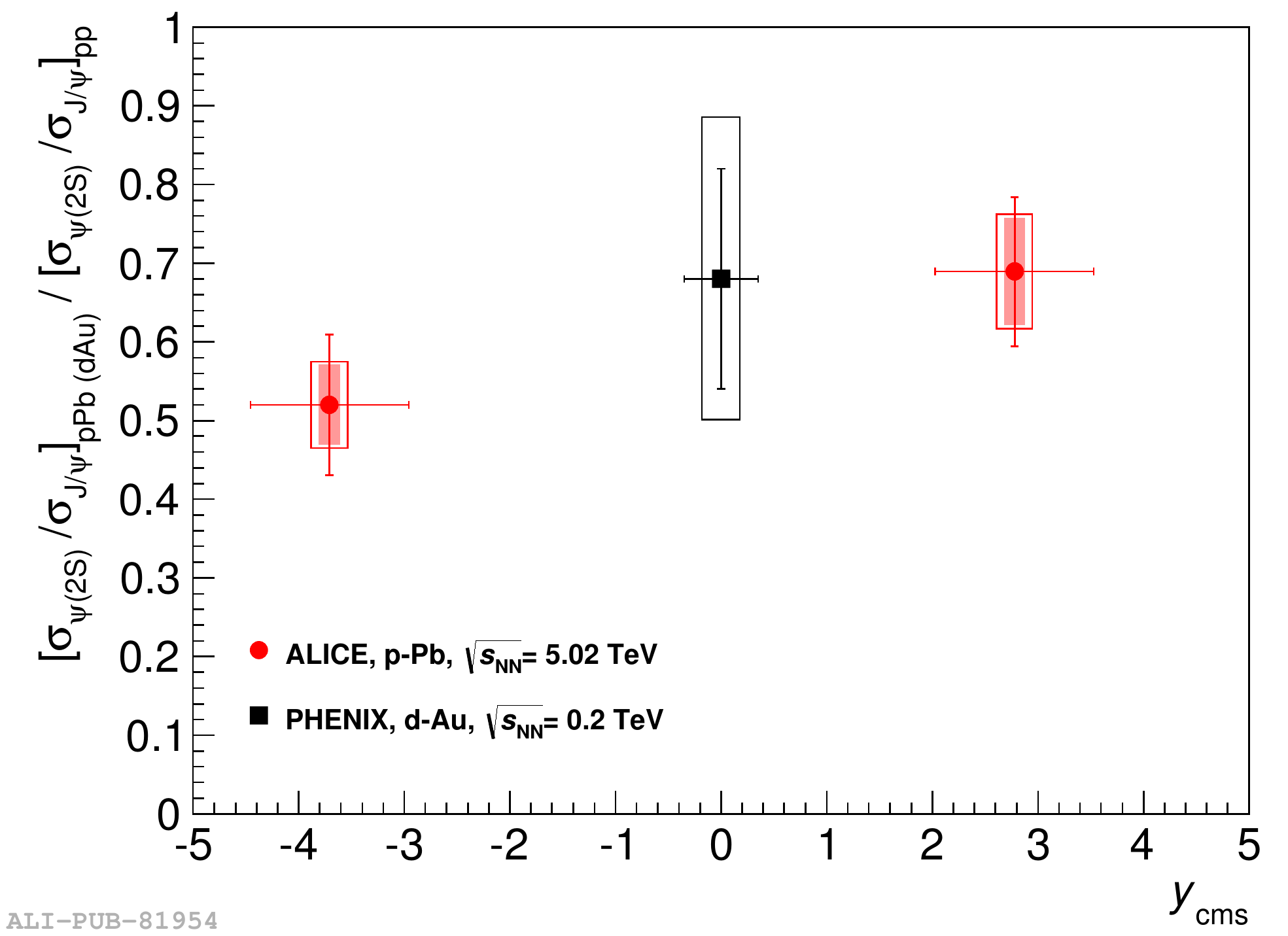}
        } 
    \end{center}
    \caption{%
Single and double $\psi$(2S)/J/$\psi$ ratio for p-Pb collisions.
     }%
   \label{fig4}
\end{figure}

In Fig.~\ref{fig4a}, we compare the B.R.$_{\rm \psi(2S)\rightarrow \mu^{+}\mu^{-}}$$\times$$\sigma_{\rm \psi(2S)}$/B.R.$_{\rm J/\psi \rightarrow \mu^{+}\mu^{-}}$$\times$$\sigma_{\rm J/\psi}$ ratio with the corresponding ALICE results in pp collisions at $\sqrt{s} = 7$ TeV (no LHC results are available at $\sqrt{s} = 5.02$ TeV). The pp ratios are significantly higher than those for p-Pb and Pb-p. The double ratio [$\sigma_{\rm \psi(2S)}/\sigma_{\rm J/\psi}]_{\rm pPb}$/[$\sigma_{\rm \psi(2S)}/\sigma_{\rm J/\psi}]_{\rm pp}$ is a useful quantity to compare the relative suppression of the two states between p-Pb and pp. In Fig.~\ref{fig4b}, the ALICE result is compared with the corresponding measurement by the PHENIX Collaboration at mid-rapidity at $\sqrt{s_{\rm NN}}$ = 200 GeV~\cite{Ada13}. Within uncertainties, a similar relative psi(2S) suppression is observed by the two experiments.

\subsection{The nuclear modification factor of $\psi$(2S)}
The nuclear modification factor of $\psi$(2S) is obtained by combining $R_{\rm pPb}^{J/\psi}$ ~\cite{R6} and the double ratio evaluated above:
\begin{center}
{$R_{\rm pPb}^{\psi(2S)} = R_{\rm pPb}^{J/\psi}\times\frac{\sigma_{\rm pPb}^{\rm \psi(2S)}}{\sigma_{\rm pPb}^{\rm J/\psi}}\times\frac{\sigma_{\rm pp}^{\rm J/\psi}}{\sigma_{\rm pp}^{\rm \psi(2S)}}$}
\end{center}

\begin{figure}[h]
     \begin{center}
        \subfigure[\hspace{0.2cm}$R_{\rm pPb}$ of J/$\psi$ and $\psi$(2S).]{%
            \label{fig5a}
            \includegraphics[width=0.49\textwidth]{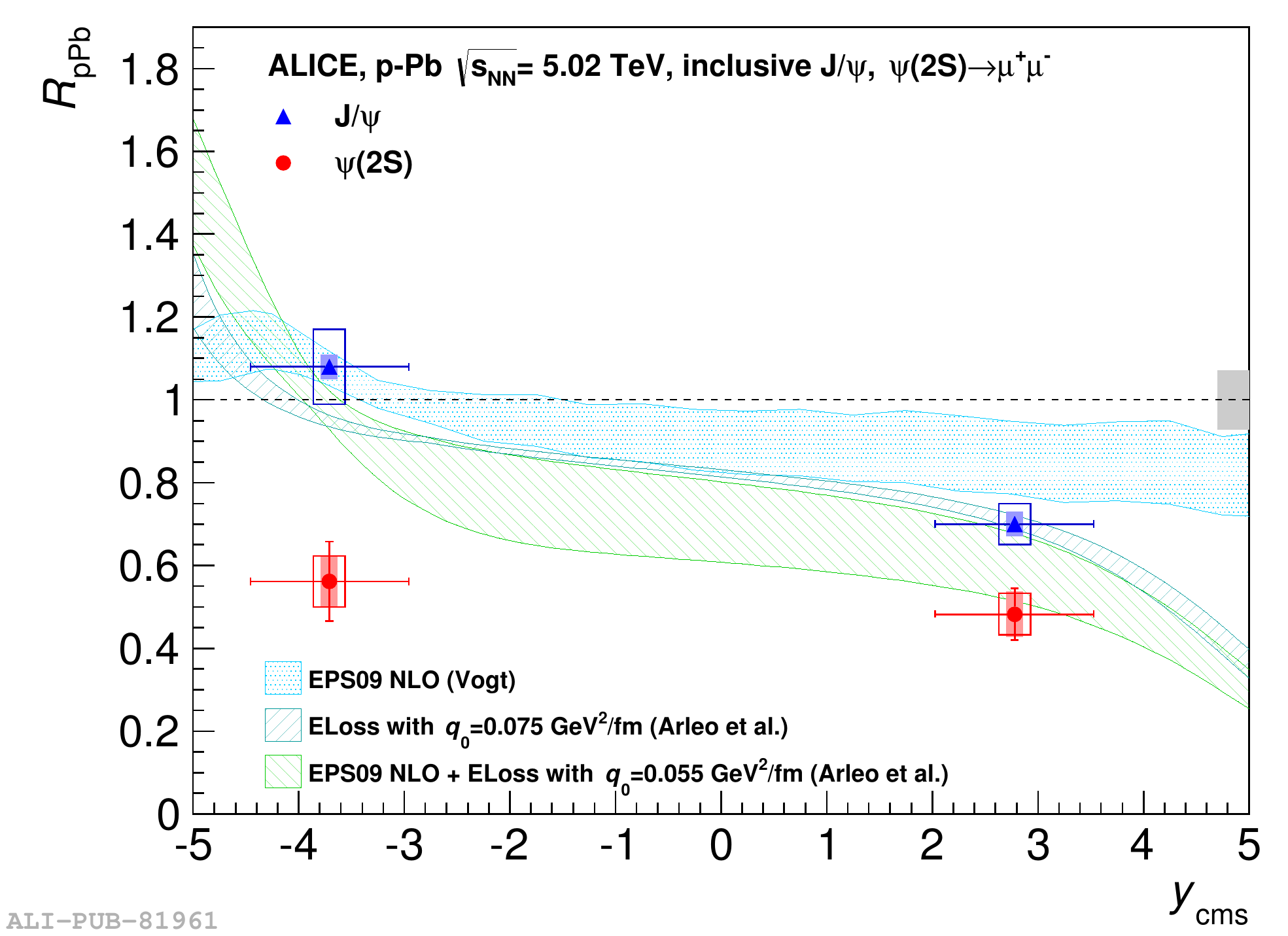}
        }%
        \subfigure[\hspace{0.2cm}$R_{\rm FB}$ of J/$\psi$ and $\psi$(2S).]{%
           \label{fig5b}
           \includegraphics[width=0.50\textwidth]{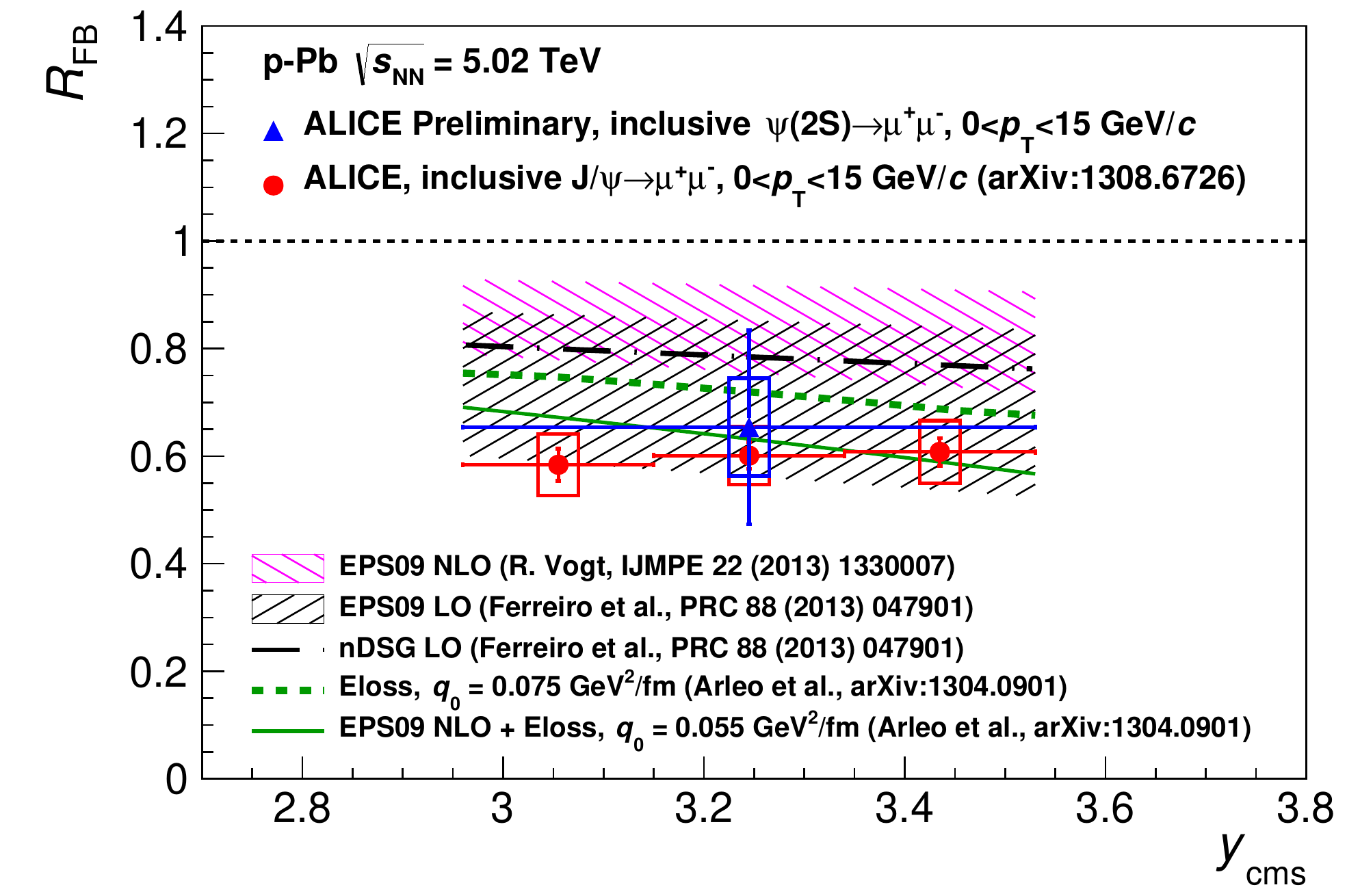}
        } 
    \end{center}
    \caption{%
$R_{\rm pPb}$ and $R_{\rm FB}$ of J/$\psi$ and $\psi$(2S) compared with various models~\cite{R9}~\cite{R7}~\cite{R8}.
     }%
   \label{fig5}
\end{figure}

In Fig.~\ref{fig5a}, $R_{\rm pPb}^{\psi(2S)}$ is compared with $R_{\rm pPb}^{J/\psi}$. The $\psi$(2S) suppression is much stronger than that of J/$\psi$ and reaches a factor $\thicksim$2 with respect to pp. The results of theoretical models, based on shadowing and energy loss, do not depend on the final quantum numbers of the charmonium states, therefore the same behaviour for J/$\psi$ and $\psi$(2S) is expected. Fig.~\ref{fig5a} shows a reasonable agreement of J/$\psi$ data with models based on EPS09 NLO shadowing calculations~\cite{R9} and on models including coherent parton energy loss~\cite{R7}, but there is a clear disagreement with $\psi$(2S) results. These results show that other mechanisms must be invoked in order to describe the $\psi$(2S) suppression in proton-nucleus collisions. More details of p-Pb analysis can be found in~\cite{R10}.

\subsection{Forward to backward ratio}
The forward to backward ratio $R_{\rm FB}$ \cite{R8} is defined in the common rapidity range 2.96 $<$ $|$\y$_{\rm cms}$$|$ $<$ 3.53 and corresponds to rapidity range 3.43 $<$ \y$_{\rm lab}$ $<$ 4.0 in p-Pb and $-$3.07 $<$ \y$_{\rm lab}$ $<$ $-$2.5 in Pb-p. This ratio does not depend on the pp interpolated cross-section at $\sqrt{s} = 5.02$ TeV and the nuclear overlap factor $T_{\rm pPb}$. Fig.~\ref{fig5b} shows the $R_{\rm FB}$ for J/$\psi$ and $\psi$(2S), compared with theoretical model calculations including shadowing and/or energy loss. All model calculations seem to be consistent with both J/$\psi$ and $\psi$(2S) results.

\section{Conclusion}
In conclusion, ALICE has measured the inclusive production cross-section of J/$\psi$ and $\psi$(2S) at forward rapidity ($2.5<y<4$) in pp collisions at a center of mass energy $\sqrt{s}=7$~TeV as a function of the quarkonium transverse momentum and rapidity. The results are in good agreement with measurements from the LHCb collaboration over similar \pt\ and \y\ ranges. 
We have also measured the inclusive $\psi$(2S) production in proton-nucleus collisions at the LHC. Measurements were performed with the ALICE Muon Spectrometer in the p-going ($2.03<y_{\rm cms}<3.53$) and Pb-going ($-4.46<y_{\rm cms}<-2.96$) directions. The $\psi$(2S)/J/$\psi$ ratio, the double ratio [$\sigma_{\rm \psi(2S)}/\sigma_{\rm J/\psi}]_{\rm pPb}$/[$\sigma_{\rm \psi(2S)}/\sigma_{\rm J/\psi}]_{\rm pp}$, $R_{\rm pPb}$ and $R_{\rm FB}$ for J/$\psi$ and $\psi$(2S), integrated over \y\ and \pt, were estimated. The results show that $\psi$(2S) is significantly more suppressed than J/$\psi$ in both rapidity regions. This observation implies that initial state nuclear effects alone cannot account for the modification of the $\psi$(2S) yields, as also confirmed by the poor agreement of the $\psi$(2S) $R_{\rm pPb}$ with models based on shadowing and/or energy loss.
Other final state effects such as $c\overline c$ interaction with the final state hadronic
medium created in p-Pb collisions should probably be considered to explain such an unexpected behaviour.


\end{document}